\documentclass{emulateapj}

\newcommand{\msun}{$M_{\odot}$}
\newcommand{\etal}{et al.~}
\newcommand{\kms}{km\,s$^{-1}$}

\shorttitle {Mild velocity dispersion evolution of  massive galaxies since
$z\sim2$} \shortauthors {A.~J.~Cenarro \& I. Trujillo}

\begin{document}

\title {Mild Velocity Dispersion Evolution of Spheroid-like Massive
  Galaxies since $z\sim2$}

\author{A. Javier Cenarro \& Ignacio Trujillo}  \affil{Instituto de
Astrof\'isica de Canarias, V\'ia L\'actea s/n, 38200 La Laguna,
Tenerife, Spain} \email{cenarro@iac.es, trujillo@iac.es}

\begin{abstract}

Making use of public spectra from Cimatti \etal(2008), we measure for
the first time the velocity dispersion of spheroid-like massive
($M_\star\sim10^{11}$\msun) galaxies at $z\sim1.6$. By comparing
with galaxies of similar stellar mass at lower redshifts, we find
evidence for a mild evolution in velocity dispersion, decreasing from
$\sim240$\,\kms\ at $z\sim1.6$ down to $\sim180$\,\kms\ at $z\sim
0$. Such mild evolution contrasts with the strong change in size (a
factor of $\sim4$) found for these type of objects in the same cosmic
time, and it is consistent with a progressive larger role, at lower
redshift, of the dark matter halo in setting the velocity dispersion
of these galaxies. We discuss the implications of our results within
the context of different scenarios proposed for the evolution of these
massive objects.

\end{abstract}


\section{Introduction} 
\label{Introduction}

Recent observations show that the most massive
($M_\star\gtrsim10^{11}$\msun) spheroid-like galaxies at $z>1.5$,
irrespective of their star formation activity (P\'erez-Gonz\'alez
\etal2008), were much smaller (a factor of $\sim4$) than their local
counterparts (Daddi \etal2005; Trujillo \etal2006, 2007; Longhetti
\etal2007; Zirm \etal2007; Toft \etal2007; Giavalisco \etal2007;
Ravindranath \etal2008, Cimatti \etal2008, hereafter C08; van Dokkum
\etal2008; Buitrago \etal2008, hereafter B08; Saracco \etal2009,
Damjanov \etal2009, Ferreras \etal2009). The near absence of such
systems ($r_{\rm e}\lesssim1.5$\,kpc; $M_\star\gtrsim10^{11}$\msun) in
the nearby Universe ($<0.03$\%; Trujillo \etal2009) implies a strong
evolution in the structural properties of these massive galaxies as
cosmic time evolves.

Different scenarios have been proposed to explain the dramatic size
evolution of these galaxies since $z\sim3$ (Khochfar \& Silk 2006;
Naab \etal2007; Fan \etal2008; Hopkins \etal2009). The main difference
among them is the role of mergers to increase the size of
galaxies. Fan \etal(2008) support an evolutionary scheme where
galaxies grow by the effect of quasar feedback, which removes huge
amounts of cold gas from the central regions hence quenching the star
formation. The removal of gas makes galaxies to puff up in an scenario
similar to the one proposed to explain the growth of globular clusters
(Hills 1980). In the merging scenario, however, merger remnants get
larger sizes than those of their progenitors by transforming the
kinetic energy of the colliding systems into potential energy. Whereas
both scenarios predict a strong size evolution for the most massive
galaxies, they disagree on the expected evolution of the velocity
dispersion of the massive galaxies at a given stellar mass. The
merging scenario basically predicts no evolution (at most a 30\% since
$z\sim3$; Hopkins \etal2009), whereas the puffing up scenario predicts
central velocity dispersions to be $\sim2$ times larger than in
present-day massive galaxies.

Constraining the evolution of the velocity dispersion of spheroid-like
massive galaxies over the last 10\,Gyr turns therefore crucial to test
the above models of galaxy evolution, as well as to determine the
importance of dark matter halos in setting the velocity dispersion of
galaxies as cosmic time evolves. For these reasons, in this Letter we
measure for the first time the velocity dispersion of such massive
galaxies at $z\sim1.6$, and compare it with a compilation of velocity
dispersions for similar galaxies at $z\lesssim1.2$. In what follows,
we adopt a cosmology of $\Omega_m=0.3$, $\Omega_\Lambda=0.7$ and
H$_0=70$\,km\,s$^{-1}$\,Mpc$^{-1}$.

\section{The Data} 
\label{TheData}

\subsection{Velocity dispersions at $z\sim1.6$}
\label{TheData16}

To determine the typical velocity dispersion of spheroid-like massive
($M_\star\sim10^{11}$\msun) galaxies at $z\sim1.6$, we have used the
publicly available stacked spectrum of galaxies at $1.4<z<2.0$
($\langle z \rangle =1.63\pm0.18$ r.m.s.~standard deviation) presented
in C08 as part of the
GMASS\footnote{http://www.arcetri.astro.it/~cimatti/gmass/gmass.html}
(Galaxy Mass Assembly ultra-deep Spectroscopic Survey) project. The
spectrum consists of an averaged spectrum of 13 massive galaxies with
a total integration time of 480\,h. Only 2 of them are classified as
disks, whereas the remaining 11 galaxies have either a pure elliptical
morphology or show a very concentrated and regular shape (see
C08). The stacked spectrum is consequently representative of massive
spheroid-like objects at that redshift. Individual galaxy spectra were
taken at VLT with FORS2, using the grism G300I and a 1\,arcsec width
slit, providing a spectral resolution of FWHM $\sim13$\,\AA\ in the
range $\lambda\lambda6000-10000$\,\AA\
($\lambda/\Delta\lambda\sim600$). Before stacking, each individual
galaxy spectrum was previously de-redshifted and assigned to have the
same weight in the $\lambda\lambda2600-3100$\,\AA\ rest-frame
range. The resulting rest-frame stacked spectrum was set to 1\,\AA/pix
over the range $\lambda\lambda2300-3886$\,\AA.

Aimed at computing the velocity dispersion of the above stacked
spectrum, it is required the use of reference template spectra of well
known spectral resolution. Section~\ref{template} provides a full
description of these data-sets, as they are basic input ingredients of
the analysis carried out in this work.

\subsection{Velocity dispersions up to $z\sim1.2$}
\label{TheData12}

We have compiled from previous work velocity dispersions estimates of
spheroid-like massive galaxies at $z<1.2$. This includes data from van
der Wel \etal(2005) ---updated, when possible, with van der Wel
\etal(2008)--- for galaxies in the ranges $0.6<z<0.8$ and $0.9<z<1.2$,
and from di Serego Alighieri \etal(2005) for galaxies at
$0.9<z<1.3$. These authors also provide stellar masses and effective
radii for their galaxies. In both samples the galaxies were classified
visually. The vast majority of these objects present a prominent
spheroidal component an are identified as either E or S0.  When
necessary, stellar masses have been transformed assuming a Chabrier
(2003) IMF. To allow a meaningful comparison along the entire redshift
interval explored in this paper, only galaxies with stellar masses in
the range $0.5<M_\star<2\times10^{11}M_\sun$ have been considered.

To have a local reference, we have retrieved velocity dispersions,
effective radii and stellar masses (Chabrier IMF) from the SDSS NYU
Value-Added Galaxy Catalog (DR6; Blanton \etal2005; Blanton \& Roweis
2007) for those galaxies within the above stellar mass range having
S\'ersic (1968) indices (in the r-band) $n>2.5$. Selecting objects
with $n>2.5$ assures that the majority of the sources have a Hubble
morphological Type T $<0$ (i.e.~ranging from E to S0; Ravindranath
\etal2004). We have estimated the average velocity dispersion of the
above galaxies in two redshift bins: $0<z<0.1$ and $0.1<z<0.2$. We
have not tried higher redshifts to assure individual signal-to-noise
ratios (S/N) larger than 10.

\section{Velocity dispersion computation} 
\label{Analysis}

\begin{deluxetable*}{ccccccc}
\tabletypesize{\scriptsize}
\tablecaption{Velocity dispersion estimates for spheroid-like massive galaxies at
$z\sim1.6$}
\tablewidth{0pt}
\tablehead{
  \colhead{Template spectra} &  
  \colhead{Spectral range (\AA) \tablenotemark{a}} & 
  \colhead{$\lambda_{\rm c}$(\AA) \tablenotemark{b}} & 
  \colhead{$\sigma_{\rm 0,templ}$ (\kms) \tablenotemark{c}}  &  
  \colhead{$\sigma_{\rm 0,C08}$ (\kms) \tablenotemark{d}} &  
  \colhead{$\Delta\sigma$ (\kms) \tablenotemark{e}} &
  \colhead{$\sigma_{\rm \star,C08}$ (\kms) \tablenotemark{f}}
 }
\startdata
CoolCAT stars        & $2510 - 3050$ & $2780$ & $ 57\pm3$ & $230$ & $267\pm20$ & $258\pm21$ \\
BC03+NGSL SSP models & $2510 - 3050$ & $2780$ & $178\pm8$ & $230$ & $178\pm23$ & $236\pm18$ \\
Keck/LRIS stars      & $3250 - 3880$ & $3565$ & $171\pm5$ & $179$ & $177\pm20$ & $236\pm15$ \\
\enddata
\tablenotetext{a}{Rest-frame spectral range employed for deriving the {\sc movel} kinematics ($\Delta\sigma$)}
\tablenotetext{b}{Central wavelength of the rest-frame spectral range}
\tablenotetext{c}{Instrumental resolution of the template spectra computed at $\lambda_{\rm c}$ (see text in Section~\ref{template}).}
\tablenotetext{d}{Instrumental resolution of the C08 stacked spectrum computed at $\lambda_{\rm c}$, using FWHM = 13\,\AA\ and $z = 1.6$.}
\tablenotetext{e}{Relative velocity dispersion with respect to that of the optimal template spectra.}
\tablenotetext{f}{Velocity dispersion of the C08 stacked spectrum of spheroid-like massive galaxies at
z$\sim1.6$, as derived from Equation.~1.}

\label{table}
\end{deluxetable*}

\subsection{The {\sc movel} program} 
\label{MOVEL}

To estimate the velocity dispersion of the C08 stacked spectrum we use
the {\sc movel} code, available at the
REDUCEME\footnote{http://www.ucm.es/info/Astrof/software/reduceme/reduceme.html}
distribution (Cardiel 1999). {\sc movel} is designed to characterize
the line-of-sight velocity distribution of a galaxy spectrum by
determining its first two moments: the mean radial velocity and the
velocity dispersion. Based on Fourier analysis, the code relies on the
{\tt MOVEL} and {\tt OPTEMA} algorithms of Gonz\'alez (1993). A full
explanation of the Fourier techniques and their particular
implementation in both routines is provided in the above work.

The {\tt MOVEL} algorithm, an improvement of the classic Fourier
quotient method by Sargent \etal(1977), is an iterative procedure in
which a galaxy model is processed in parallel to the galaxy
spectrum. In this way, a comparison between the input and recovered
broadening functions for the model allows to correct the galaxy power
spectrum from any imperfections of the data handling in Fourier
space. The {\tt OPTEMA} algorithm is designed to overcome the typical
template mismatch problem by computing an optimal template for the
galaxy spectrum. It is fed with a set of representative template
spectra which are scaled, shifted and broadened according to initial
values for the mean line strength, the radial velocity, and the
velocity dispersion. Then, the algorithm finds the linear combination
of the template spectra that best matches the observed galaxy
spectrum. This constitutes a first composite template which is fed
into the {\tt MOVEL} algorithm. The output kinematic parameters are
then used to create an improved composite template and the process is
iterated until it converges. This iterative approach then provides an
optimal template while simultaneously computing the radial velocity
and velocity dispersion of the galaxy spectrum.

Ideally, the reference template spectra are chosen to be
representative stellar spectra observed with an identical technical
setup to that of the galaxy spectra. This way, the instrumental
broadening is the same in both data sets, and the broadening
difference between galaxy and template spectra, $\Delta\sigma$, is a
direct measurement of the galaxy starlight velocity dispersion,
$\sigma_\star$. Since this is not the case for the C08 stacked
spectrum, the different instrumental resolutions of the distinct data
sets employed in this work must be taken into account. This is
explained in detail in Sections~\ref{template} and~\ref{results}.

\subsection{The template spectra}
\label{template}

We have restricted the kinematical analysis of the C08 spectrum to
those regions containing strong spectral features from the galaxy
starlight. Because of the lack of available template spectra of
high-enough spectral resolution covering the full spectral range of
interest, we employ three different data sets. Two of them allow us to
analyze the near ultraviolet (NUV) region at $\lambda\leq3050$\,\AA,
which comprises e.g.~several Fe{\sc ii} lines at
$\lambda\sim2600$\,\AA, the $\lambda2800$ Mg{\sc ii} doublet and the
$\lambda2852$ Mg{\sc i} line. The third template set is chosen to
provide a complementary analysis from the region
$\lambda\geq3250$\,\AA, with features such as the $\lambda3360$ NH
and $\lambda3862$ CNO bands. The three sets of template spectra,
whose properties are summarized in Table~\ref{table}, are described
below:

i) First, from the
CoolCAT\footnote{http://casa.colorado.edu/~ayres/CoolCAT/} database,
an HST/STIS echelle catalog of normal late-type stars with R
$\sim40000$ (Ayres 2005), we have selected eight spectra (from F5V to
G0III spectral types) covering the range
$\lambda\lambda2510-3050$\,\AA\ with no gaps in between. They have
been re-binned at 1\,\AA/pix to match the linear dispersion of the C08
spectrum. Using {\sc movel} and their corresponding high-resolution
CoolCAT spectra as reference templates, we have derived that the
spectral resolution of the re-binned templates is $\sigma_{\rm
STIS}\sim57\pm3$\,\kms\ at $\lambda\sim2800$\,\AA, which corresponds
to R $\sim2240$.

ii) The second set of template spectra consists of a subsample of
nineteen SSP spectral energy distributions by G.~Bruzual (private
communication). They are based on the Bruzual \& Charlot (2003) code
and the Next Generation Spectral Library, NGSL (Gregg \etal2004; Heap
\& Lindler 2007), with an average resolution of R $\sim1000$. The
employed SSP models span ages from 0.3 to $3$\,Gyr with metallicities
[Z/H] $= -0.4$, $0.0$, and $+0.4$, hence enclosing the 1\,Gyr and
solar metallicity estimate of C08 for their stacked spectrum. Again,
using {\sc movel} and the above high-resolution CoolCAT spectra, we
have accurately derived the spectral resolution of Bruzual's models in
the range $\lambda\lambda2510-3050$\,\AA, which is $\sigma_{\rm
BC03+NGSL}\sim178\pm8$\,\kms, or equivalently R $\sim710$.

iii) Finally, for the red side of the C08 stacked spectrum, we employ
a set of eighteen stellar spectra (dwarfs and giants ranging from G0
to K7 spectral types) observed with LRIS at Keck as radial velocity
templates for a different program (Beasley \etal2009). Having used the
grism 600/4000 and the 1\,arcsec longslit, the stellar spectra have a
FWHM of $4.77\pm0.14$\,\AA\ ($\sigma_{\rm LRIS}\sim171\pm5$\,\kms\ at
3565\,\AA; R $\sim750$) as determined from the width of the wavelength
calibration arc lines. The spectra span the range
$\lambda\lambda3150-5600$\,\AA, although only the blue side is
employed here.

\subsection{Results and robustness of the method}
\label{results}

We have analyzed the stacked spectrum of C08 with {\sc movel} using
the three sets of template spectra described in
Section~\ref{template}. In each case, to constrain the reliability of
the final results, the procedure has been repeated several times using
different initial {\sc movel} parameters, like e.g.~different starting
velocity dispersions (from 50 up to 300\,\kms, in steps of 50\,\kms).

For each single {\sc movel} solution, the stellar velocity dispersion
of the C08 stacked spectrum, $\sigma_{{\star},{\rm C08}}$, is derived
as

\begin{equation}
\sigma_{{\star},{\rm C08}} = \sqrt{
\Delta\sigma ^{2} + \sigma_{\rm 0,templ}^{2} -
\left( \begin{array}{c} \frac{\sigma_{\rm 0,C08}}{1+z}\end{array} \right) ^{2}, 
}
\end{equation}
where $\Delta\sigma$ is the broadening difference between the optimal
template spectra and the C08 spectrum (as derived from {\sc movel}),
$\sigma_{\rm 0,templ}$ and $\sigma_{\rm 0,C08}$ are their intrinsic
instrumental resolutions, and $z$ is the redshift ($\sim1.6$). Note
that, since the individual galaxy spectra of C08 were previously
blue-shifted before stacking, the effective instrumental resolution at
rest-frame decreases by ($1+z$).

For each set of template spectra, Table~\ref{table} presents
$\sigma_{\rm 0,templ}$, $\sigma_{\rm 0,C08}$, and the mean values of
$\Delta\sigma$ and $\sigma_{{\star},{\rm C08}}$, with their errors
accounting for the r.m.s.~standard deviation of the solutions obtained
from the different {\sc movel} trials. An error-weighted mean of the
three independent $\sigma_{{\star},{\rm C08}}$ solutions, which are
indeed consistent within the uncertainties, gives
$\langle\sigma_{{\star},{\rm C08}}\rangle =241\pm10$\,\kms. 

It is worth noting that the above value could be slightly
overestimated, as redshift uncertainties for the 13 stacked galaxy
spectra may have enlarged artificially the width of the lines. We have
also checked that the strong $\lambda2800$ Mg{\sc ii} doublet in the
stacked spectrum ($\sim8$\,\AA, Table~2 of C08) is not significantly
broadened by potential non-stellar Mg{\sc ii} absorptions in
surrounding galactic gas clouds. Such absorptions are typically much
narrower ($\sim45$\,\kms, as they follow much colder kinematics) and
$\sim5$ times weaker ($\sim1.6$\,\AA, for typically $\sim14$ clouds;
see e.g.~Steidel \etal1994; Churchill \etal2000). Therefore, even if
they were present in all galaxies, $\langle\sigma_{{\star},{\rm
C08}}\rangle$ could be overestimated by only $\sim1$\,\kms\ (up to
$\sim14$\,\kms\ if all galaxies were Damped-Ly$\alpha$ systems, which
is very unlikely as no Lyman absorptions are detected). On the other
hand, if the two disk galaxies in the stacked spectrum had lower
typical velocity dispersions of e.g.\,$150$\,\kms, the derived
$\langle\sigma_{{\star},{\rm C08}}\rangle$ would underestimate the
true velocity dispersion of early-type galaxies by
$\sim13$\,\kms. Overall, since the above systematics may cancel each
other, $241\pm10$\,\kms is probably a reasonable estimation of the
velocity dispersion of $M_\star\sim10^{11}$\msun\ spheroid-like
galaxies at $z\sim1.6$.

\begin{figure}
\epsscale{1.1}
\plotone{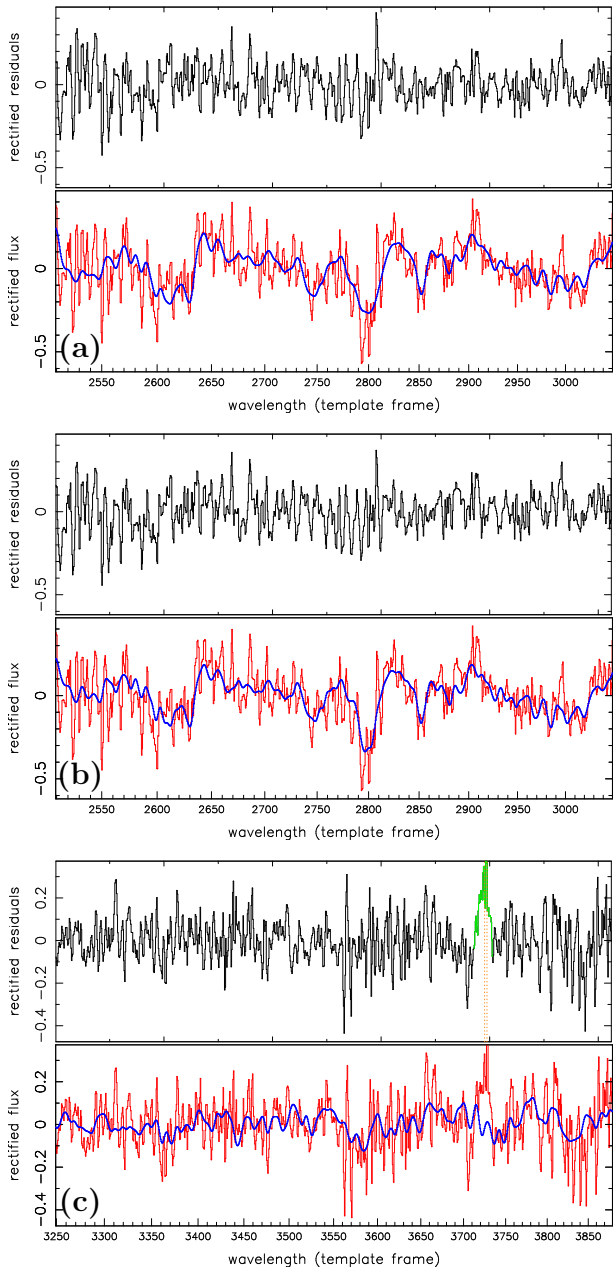}
\caption{Representative examples of the {\sc movel} fitting results
  derived for the C08 stacked spectrum and the three sets of template
  spectra: (a) CoolCAT stars, (b) BC03$+$NGSL SSP models, and (c)
  Keck/LRIS stars. In each case, bottom panels illustrate the C08
  stacked spectrum of galaxies at $z\sim1.6$ (red) and the optimal
  template spectrum (blue, thick line) broadened at the mean
  $\Delta\sigma$ values in Table~\ref{table}. Top panels show the
  residuals of the fits. In panel $c$, the region around the O\,{\sc
  ii} emission line doublet at $\lambda3727$\,\AA\ (dotted lines) was
  masked and rejected from the C08 stacked spectrum (green) for the
  fitting computation.}
\label{fig1}
\end{figure}

Figure~\ref{fig1} illustrates representative {\sc movel} solutions as
derived from each template set. At the NUV, CoolCAT stars (panel $a$)
provide a slightly more broadened solution than the NGSL--based SSP
models (panel $b$), probably due to the intrinsic differences between
the two template sets and their effects on the template mismatch
problem.  At longer wavelengths (panel $c$), the galaxy spectral
features look somewhat blurred, probably by sky line residuals
commonplace of the near-infrared region at observed frame. Still, in
all cases the solutions reproduce nicely the C08 spectrum.

To further check the reliability of the {\sc movel} results, we
carried out Monte Carlo simulations in which we added Gaussian noise
(to match S/N = 5 to 30\,\AA$^{-1}$) to artificially broadened
reference template spectra ($\Delta\sigma=100-400$\,\kms, in steps of
50\,\kms) from our three template sets. For such {\it fake} galaxy
spectra, we run the {\sc movel} procedure in the same way as we did
for the C08 stacked spectrum. As expected, the accuracy in the final
solution decreases with the decreasing S/N of the problem
spectrum. The typical scatter of the derived $\Delta\sigma$ solutions
---averaged over all the simulations and initial {\sc movel}
conditions--- are 55, 24, 12, and 8\,\kms\ for S/N = 5, 10, 20, and
30\,\AA$^{-1}$ respectively. More importantly, for all S/N values, the
recovered mean $\Delta\sigma$ is statistically consistent with the
assumed input value, hence supporting the reliability of both the
method and $\sigma_{{\star},{\rm C08}}$.

\section{Discussion} 
\label{ConclusionsandDiscussion}

The fact that the velocity dispersion of $M_\star\sim10^{11}$\msun\
spheroid-like galaxies at $z\sim1.6$ is $\sim240$\,\kms has important
consequences to understand their evolution:

\begin{itemize} 

\item It confirms that high-z spheroid-like massive galaxies are truly
massive objects. The unexpected strong size evolution of these objects
has cast some doubts about the reliability of their stellar mass
estimates, which relies on the assumption that the IMF is the same at
all redshifts. In fact, an appropriate change of the IMF with redshift
would mitigate the problem of the strong size evolution. However, the
derived velocity dispersion at $z\sim1.6$ is similar to that of
present-day galaxies with $M_\star>10^{11}$\msun, consequently
constraining any potential change of the IMF with redshift.

\item It alleviates the problem of understanding how massive ---and
compact--- galaxies at high-z can evolve through merging since that
epoch. An extraordinarily high velocity dispersion at high-z
(e.g.~$\sim400$\,\kms) would have implied that the gravitational
potential depth of the system would be so intense that it would not
easily evolve in size.

\end{itemize}

\begin{figure}
\epsscale{1.15}
\plotone{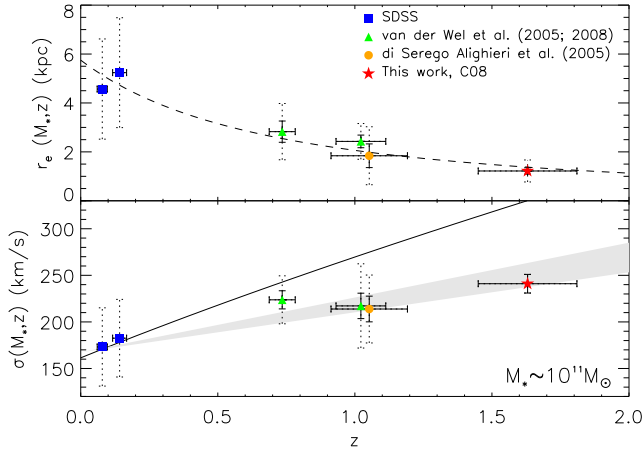}
\caption{\textit{Top Panel:} Size evolution of
  $M_\star\sim10^{11}M_{\sun}$ spheroid-like galaxies as a function of
  redshift. Different symbols show the median values of the effective
  radii for the different galaxy sets considered in this work (see
  Section~\ref{TheData}), as indicated in the labels. Dashed error
  bars, if available, show the dispersion of the sample, whereas the
  solid error bars indicate the uncertainty of the median value. The
  dashed line represents the observed evolution of sizes $r_{\rm
  e}(z)\propto(1+z)^{-1.48}$ found in B08 for galaxies of similar
  stellar mass. \textit{Bottom Panel:} Velocity dispersion evolution
  of the spheroid-like galaxies as a function of redshift, with
  symbols as given above. Assuming the B08 size evolution, the solid
  line represents the prediction from the "puffing up" scenario (Fan
  \etal2008), whereas the grey area illustrates the velocity
  dispersion evolution within the merger scenario of Hopkins
  \etal(2009) for $1<\gamma<2$. See text for details. }
\label{fig2}
\end{figure}

To put in context the above result, Figure~\ref{fig2} shows the sizes
(top panel) and velocity dispersions (bottom panel) for the C08
galaxies and the compilation of spheroid-like galaxies of similar mass
described in Section~\ref{TheData}. These galaxies follow nicely the
observed strong size evolution found in other independent larger
samples where completeness effects are accounted (Trujillo et
al. 2007; B08; dashed line), hence probing that the sizes of the
objects explored in this paper are typical of average spheroidal-like
massive galaxies at those redshifts. It is also clear that, although
massive spheroid-like galaxies have experienced a strong size
evolution (a factor of $\sim4$ since $z\sim1.5$; Trujillo \etal2007),
the velocity dispersion has evolved mildly by a factor of $\sim1.3$ in
the same redshift interval (see Bernardi 2009 for a similar trend at
$z<0.3$). This result is crucial to constrain the different scenarios
proposed so far to explain the dramatic size evolution since $z\sim2$,
as they disagree in the amount of evolution expected in their velocity
dispersions.

In the puffing up scenario of Fan \etal(2008), velocity dispersions
change as $\sigma_\star\propto r_{\rm e}^{-1/2}$. Using the observed
size evolution of B08, $r_{\rm e}(z)\propto (1+z)^{-1.48}$, the solid
line in Fig.~\ref{fig2} illustrates the expected $\sigma_\star$
evolution under this scenario. It increases with redshift in a way
that galaxies are expected to double their $\sigma_\star$ at
$z\sim1.5$. The observed velocity dispersions are compatible ---within
the error bars--- with this scenario up to $z\sim0.7$. However, at
$z>1$ the discrepancy between theory and data is apparent.

In the merging scenario of Hopkins \etal(2009), at a fixed stellar
mass, $\sigma_\star$ evolves with redshift as
\begin{equation}
\frac{\sigma_{\star}(z)}{\sigma_{\star}(0)}\propto\frac{1}{\sqrt{1+\gamma}}\sqrt{\gamma+\frac{r_{\rm
e}(0)}{r_{\rm e}(z)}},
\end{equation}
where $\gamma\equiv(M_{\rm halo}/R_{\rm halo})/(M_{\star}/r_{\rm
  e}$) is the dark matter contribution to the central potential
relative to that of the baryonic matter at $z\sim0$, and $R_{\rm
  halo}$ is the effective radius of the halo. Again, we have used the
B08 size evolution as an input into the merging model prediction. The
grey area in Fig.~\ref{fig2} encloses the expected $\sigma_\star$
evolution when $\gamma$ varies between 1 and 2. The agreement between
the merging scheme and the observed evolution looks reasonably good at
all redshifts.

Following Hopkins \etal(2009), to explain why $\sigma_\star$ has only
changed weakly since $z\sim2$ it is necessary to consider that the
observed velocity dispersion is driven by two components: the baryonic
matter and the dark matter halo. Both components contribute linearly
to the central gravitational potential of the galaxy, hence
$\sigma_\star\propto (M_{\star}/r_{\rm e}+M_{\rm halo}/R_{\rm
halo}$). Assuming that $R_{\rm halo}$ evolves weakly with time (most
simulations show that halos build inside-out, so the central potential
is set first and just the outer halo grows with time), the dark matter
effect on the central potential of the galaxy (i.e.~on $\sigma_\star$)
basically remains unchanged at a fixed $M_{\star}$. However, the
influence of the baryonic matter on the gravitational potential has
changed strikingly since $z\sim1.5$, at present-day being $\sim4$
times smaller due to the expansion of the object. The relative
influence of the dark matter on setting the inner potential increases
with the decreasing effect of the baryonic matter.  In fact, the data
look in agreement with an almost symmetric influence of dark and
baryonic matter (i.e.~$\gamma\sim1$) in setting the central
gravitational potential of present-day objects.

\acknowledgments 

We acknowledge the referee for constructive comments. This work
has been possible thanks to the helpful contribution of several
people. G.~Bruzual provided us with his SSP model predictions based on
the NGSL. Fruitful discussions with P.~S\'anchez-Bl\'azquez,
A.~Vazdekis and C.~Conselice helped us to find the right strategy to
achieve our goals. N.~Cardiel gave us very valuable inputs on the {\sc
movel} capabilities. Finally, we benefited from discussion with
R.~Guzm\'an, M.~Balcells, M.~L\'opez-Corredoira, and A.~van der
Wel. A.J.C. and I.T.C. are Juan de la Cierva and Ram\'on y Cajal
Fellows of the Spanish Ministry of Science and Innovation. This work
has been funded by the Spanish Ministry of Science and Innovation
through grant AYA2007-67752-C03-01.

{}

\end{document}